\documentclass{article}
\usepackage{graphicx}
\usepackage{epsfig}
\usepackage{amsmath}
\usepackage{amssymb}
\def\bea{\begin{eqnarray}}
\def\eea{\end{eqnarray}}

\begin{document}

\baselineskip=16pt
\begin{titlepage}
\setcounter{page}{0}
\begin{center}

\vspace{0.5cm}
 {\Large \bf Dbrane Near NS5-branes: with Electromagnetic Field}\\
\vspace{10mm} {\large Bin Chen$^{1}$\footnote{e-mail
 address:bchen@itp.ac.cn},\hspace{2ex}Miao Li$^{1,2}$\footnote{e-mail address: mli@itp.ac.cn} and
 Bo Sun$^{2,3}$\footnote{e-mail address: sunbo@itp.ac.cn}}\\
\vspace{12mm} {\it $~^1$ Interdisciplinary Center of
Theoretical Studies, Chinese Academy of Science,\\
 P.O. Box 2735, Beijing 100080, P.R. China}\\

 \vspace{6mm} {\it $~^2$
      Institute of Theoretical Physics, Chinese Academy of
      Sciences,\\
       P.O. Box 2735,
      Beijing 100080, P.R. China\\}

 \vspace{6mm} {\it $~^3$ Graduate School of the Chinese Academy of
 Science, Beijing 100080, P.R. China\\}

\vspace*{5mm} \normalsize
\smallskip
\medskip
\smallskip
\end{center}
\vskip0.6in \centerline{\large\bf Abstract} \vspace{1cm} {We study
the Dp-brane dynamics near NS5-branes with constant
electromagnetic field. In the framework of effective
Dirac-Born-Infeld action, we investigate the effect of the
electromagnetic field on the Dp-brane dynamics. The radial motion
of the Dp-brane on the transverse directions of NS5-branes can be
mapped to a rolling tachyon in a constant NS $B_{\mu\nu}$
background. In the near throat region, the classical motion can be
identified with the rescaled hairpin. After constructing the
boundary state of the rescaled hairpin, we discuss the closed
string emission of the Dp-brane and find that the energy of the
closed string emission is always finite in the presence of
constant electric field. Taking the winding strings into account,
the emitted energy is divergent, indicating that the emitted
winding strings carry away most of the energy. }

\vspace*{2mm}

\end{titlepage}

\section{Introduction}

The subject of time-dependent solutions in string theory is
important since its study has been expected to provide us with the
knowledge about the resolution of the black hole singularity and
the cosmological singularity. The rolling tachyon, being an
exactly solvable time-dependent model in the classical open string
theory, has drawn much attention since proposed by A. Sen
\cite{Sen1, Strominger1}. The condensation of  the rolling tachyon
leads to closed string radiation and the relics of the closed
string emission may be identified with the tachyon matter
\cite{Chen, Okuda, Liu}, which is a pressureless fluid.
Nevertheless, the closed string emission remains an intriguing
issue. Another remarkable fact is that the low energy effective
action actually describes many aspects of the rolling tachyon
condensation quite well \cite{Sen2}. The origin of this agreement
has been investigated in \cite{Kutasov1}.

In the study of the rolling tachyon, a very interesting
generalization is to implement a constant NS $B_{\mu\nu}$
background, or a constant electromagnetic flux, on the worldvolume
of the unstable Dp-brane. In the case of the rolling tachyon with
constant flux, the DBI action is still a good description of the
dynamics of the rolling tachyon \cite{Sen3,Rey}. It turns out that
in the end of the tachyon condensation there is not only
pressureless tachyon matter but also string fluid induced by the
electric field \cite{Yi1}. It was also found that the presence of
electric field can slow down the emission of closed strings. In
particular, it has been shown that the emitted energy of closed
strings without winding is finite and negligible \cite{Nagami}.
However, after taking into account the winding strings, the
emitted energy is divergent when $p\leq 3$, suggesting that most
of the emitted energy goes to the highly wound closed strings
\cite{Yi2}.

Recently, D. Kutasov noticed that the behavior of Dp-brane near
NS5-branes is similar to the rolling tachyon \cite{Kutasov2}. Let
the world-volume of NS5-branes lie along $x^0,\cdots x^5$ and the
one of Dp-brane lie along $x^0,\cdots x^p$ with $p \le 5$. In both
type II string theories, such a configuration breaks supersymmetry
completely and is unstable. It was argued that the motion of
Dp-brane can be described reliably in a wide range by the DBI
action in the supergravity background of NS5-branes. The radial
motion of the Dp-brane in the transverse direction $x^6,\cdots
x^9$ can be mapped to the motion of the rolling tachyon after
field redefinition. When the Dp-brane enters the deep throat
region of NS5-branes, it behaves like a supersymmetric hair-pin
brane, which can be described by a boundary state of corresponding
conformal field theory in the throat. The conformal field theory
turns out to be $R^{1,5}\times SU(2)_k \times R_\phi$, where
$SU(2)_k$ is the three-sphere with radius $\sqrt{k}$ and $R_\phi$
is the CFT of a linear dilaton \cite{CHS}. The boundary state of
the hair-pin brane was constructed in \cite{Nakayama}.  The closed
string emission of the Dp-brane near NS5-branes was discussed in
\cite{Sahakyan, Nakayama}. Other related discussions can be found
in \cite{Rey2}.

In this paper, we will study the Dp-brane dynamics near NS5-branes
with an electromagnetic field along the worldvolume of Dp-brane.
Here we will use the effective DBI description to study various
aspects of Dp-brane dynamics. Using T-duality approach, we manage
to construct the boundary state of the modified hairpin brane and
study the closed string emission with and without the winding
strings.

In section 2, we discuss the motion of Dp-brane using the DBI
action. In section 3, we will construct the boundary state of the
modified hairpin brane. And in section 4 and 5, we discuss the
closed string emission without and with the winding strings
respectively. We conclude the paper in section 6.


\section{DBI Analysis}

In this section we use the effective action (DBI action) for the
Dp-brane to analyze its dynamics near NS5-branes. The tension of a
NS5-brane scales as $1/g_s^2$ while the tension of a $Dp$-brane
scales as $ 1/g_s$, so it is natural to take NS5-branes'
supergravity solution as a background for the Dp-brane when the
string coupling is weak.

The coordinates on the world-volume of $k$ coincident NS5-branes
are $x^\mu$, $\mu=0,1,...5$, and we use $x^n,n=6,7,8,9$ to label
the four transverse dimensions. Let the Dp-brane be parallel to
NS5-branes and let the worldvolume of Dp-brane lie along $x^0,
\cdots x^p$ with $2 \le p \le 5$. Such a system breaks
supersymmetry completely and is unstable.

Setting $r^2=\displaystyle{\sum_{n=6}^{9} x^nx^n}$, the low energy
supergravity solution of NS5-branes is
\begin{eqnarray}\label{metric}
ds^2&=&dx^\mu dx_\mu+H(r)dx^ndx_n\equiv g_{MN}dx^Mdx^N\nonumber\\
\frac{g_s(\Phi)}{g_s}&=&\exp(\Phi-\Phi_0)=\sqrt{H(r)}\nonumber\\
H_{mnp}&=&-\epsilon_{qmnp}\partial^q \Phi,\nonumber \\
H(r)&=&1+\frac{kl_s^2}{r^2},
\end {eqnarray}
where $H_{mnp}$ is the NS 2-form field strength, $g_s$ is the
asymptotic string coupling, and $l_s$ is the string length unit.

Since there is an $SO(4)$ rotational symmetry for the four
transverse dimensions, without loss of generality we focus on the
radial motion of a Dp-brane.  The dynamics of Dp-brane is well
 described by the DBI action
\begin{equation}\label{action}
S_p=-\tau_p\int d^{p+1}x
e^{-(\Phi-\Phi_0)}\sqrt{-det(G_{\mu\nu}+B_{\mu\nu})}
\end{equation}
where $\tau_p$ is the asymptotic tension of the Dp-brane
\begin{equation}\label{tau}
\tau_p\sim\frac{1}{g_sl_s^{p+1}}
\end{equation}
Now we turn on the electromagnetic field
$B_{01}=e,B_{10}=-e,B_{12}=b,B_{21}=-b$. After pull back to the
worldvolume of the Dbrane, it reads\footnote{Here we take $p=3$,
it is straightforward to generalize to other cases.}
\begin{equation}
\mathbf{G_{\mu\nu}}= \left(\begin{array}{cccc}
-1+H\dot{r}^2&0&0&0\\
0&1&0&0\\
0&0&1&0\\
0&0&0&1\\
\end{array}\right)\hspace{5ex}
\mathbf{B_{\mu\nu}}= \left(\begin{array}{cccc}
0&e&0&0\\
-e&0&b&0\\
0&-b&0&0\\
0&0&0&0\\
\end{array}\right).
\end {equation}
The radial position of the Dbrane gives rise to a scalar field
$r(x^\mu)$ on the worldvolume. We are only interested in the case
when $r$ is just a function of
 $t$, so the action is
\begin{eqnarray}
S_p&=&-\tau_pV\int
dt\frac{1}{\sqrt{H(r)}}\sqrt{(1+b^2)(1-H\dot{r}^2)-e^2}\nonumber\\
&=&-\tau_pV\int
dt\sqrt{\frac{1}{H(r)}(1+b^2-e^2)-\dot{r}^2(1+b^2)}.
\end{eqnarray}
Similar to the case without the electromagnetic flux, by defining
a ``tachyon" field $\frac{dT}{dr}=\sqrt{H(r)}$ one can map the
above action to the DBI action of the rolling tachyon in the
presence of constant electromagnetic field but with tachyon
potential $V(T)=\frac{\tau_p}{\sqrt{H(r)}}$. Since we are mainly
interested in the near throat region where CHS theory\cite{CHS}
works, a rescaled parameter is more convenient: take $R=g_s^{-1}r$
to be finite
 while $g_s\rightarrow0,
r\rightarrow0$. In this limit $H(r)$ is simply $H(r)\sim r^{-2}$,
thus the action can be written as:
\begin{eqnarray}
S_p&=&-\tau_pV\int
dt\sqrt{\frac{g_s^2(1+b^2-e^2)}{kl_s^2}R^2-g_s^2(1+b^2)\dot{R}^2}\nonumber\\
&=&-T_pV\int
dt\frac{R}{\sqrt{k}l_s}\sqrt{(1+b^2-e^2)-(\frac{d}{dt}\log
R)^2(1+b^2)kl_s^2}
\end{eqnarray}
where $T_p\equiv\tau_pg_s$. Recall (\ref{tau}), the effective tension $T_p$ does not vanish in this limit. \\
In terms of the rescaled radial coordinate, the solution reads:
\begin{eqnarray}
R=\frac{\sqrt{k}l_s}{A\cosh\frac{\sqrt{1+b^2-e^2}t}{l_s\sqrt{k(1+b^2)}}}
\end{eqnarray}
where A is an integration constant depending on the initial
condition and is related to the conserved energy tensor $\rho$ as
in (\ref{decaytensor}), $A={T_p(1+b^2)}/{\rho\sqrt{1+b^2-e^2}}$.
The linear dilaton $\Phi = -\phi/\sqrt{k}l_s$, where $\phi$ is the
proper distance, and \bea\label{EOM}
e^{-\frac{\phi}{\sqrt{k}l_s}}=A\cosh\frac{\sqrt{1+b^2-e^2}t}{l_s\sqrt{k(1+b^2)}}
\eea In late time, $t\rightarrow \infty$, one has \bea \phi \sim
-\gamma^{-1}|t|, \eea with \bea
\gamma^{-1}=\frac{\sqrt{1+b^2-e^2}}{\sqrt{(1+b^2)}} < 1. \eea So
the Dbrane approaches a limiting speed $\gamma^{-1}$ in the late
time. It is due to the presence of the electromagnetic field that
the limiting speed is less than the speed of light.

It is remarkable that (\ref{EOM}) is a modification of hairpin
brane discussed in \cite{Nakayama, Sahakyan}. We will show in the
next section, by using T-duality approach, that one can construct the
boundary state for this brane and then discuss the closed string
emission issue.

It is easy to calculate the energy-momentum tensor. From
$(\ref{action})$, one has:
\begin{eqnarray}
\delta S&=&-\frac{\tau_p}{2}e^{-(\Phi-\Phi_0)}\sqrt{-det(\mathbf{G}+\mathbf{B})}(\mathbf{G}+\mathbf{B})^{\mu\nu}(\delta g_{\mu\nu}+\delta b_{\mu\nu})\nonumber\\
&=&-\frac{\tau_p}{2}\frac{1}{\sqrt{H(r)}}\sqrt{-det(\mathbf{G}+\mathbf{B})}(\mathbf{G}+\mathbf{B})^{\mu\nu}(\delta
g_{\mu\nu}+\delta b_{\mu\nu})\nonumber
\end{eqnarray}
We get the energy-momentum tensor $T^{\mu\nu}$ and NS source
$S^{\mu\nu}$. For the above solution (\ref{EOM}), the
non-vanishing components are:
\begin{eqnarray}\label{decaytensor}
T^{00}&=&\frac{T_p(1+b^2)}{A\sqrt{1+b^2-e^2}}=\rho\nonumber\\
T^{11}&=&\frac{\rho}{1+b^2}(\frac{(1+b^2-e^2)\tanh^2(\frac{\sqrt{1+b^2-e^2}\;t}{l_s\sqrt{k(1+b^2)}})}{1+b^2}-1)\nonumber\\
T^{22}&=&\frac{\rho}{1+b^2}(\frac{(1+b^2-e^2)\tanh^2(\frac{\sqrt{1+b^2-e^2}\;t}{l_s\sqrt{k(1+b^2)}})}{1+b^2}-1+e^2)\nonumber\\
T^{02}&=&\frac{eb\rho}{1+b^2}\nonumber\\
T^{33}&=&\frac{\rho}{1+b^2}(1+b^2-e^2)[\tanh^2(\frac{\sqrt{1+b^2-e^2}\;t}{l_s\sqrt{k(1+b^2)}})-1]=T_s\nonumber\\
S^{02}&=&\frac{-e\rho}{1+b^2}\nonumber\\
S^{12}&=&\frac{-b\rho}{1+b^2}(\frac{(1+b^2-e^2)\tanh^2(\frac{\sqrt{1+b^2-e^2}\;t}{l_s\sqrt{k(1+b^2)}})}{1+b^2}-1)
\end{eqnarray}
The above result reduces to the expressions in \cite{Kutasov2,
Sahakyan} in the absence of an electromagnetic field. The
off-diagonal elements of $T^{\mu\nu}$ can be recognized as the
Poyntin tensor since we have orthogonal electric and magnetic
field. In fact, the constant electromagnetic field background can
be thought of being generated from an ideal liquid of fundamental
strings, each stretching along ${x^0,x^1}$ and rigid flowing along
${x^2}$. Boosting to the rest frame of fundamental strings will
make $b=0$, diagonalizing $T^{\mu\nu}$. Taking
$t\rightarrow\infty$, the late-time behavior is $R\rightarrow 0$
and:
\begin{eqnarray}\label{fstringtensor}
T^{00}&=&\rho\nonumber\\
T^{11}_\infty&=&\frac{-\rho e^2}{(1+b^2)^2}\nonumber\\
T^{22}_\infty&=&\frac{\rho e^2b^2}{(1+b^2)^2}\nonumber\\
T^{02}_\infty&=&\frac{eb\rho}{1+b^2}\nonumber\\
T^{33}_\infty&=&0\nonumber\\
S^{02}_\infty&=&\frac{-e\rho}{1+b^2}\nonumber\\
S^{12}_\infty&=&\frac{-be^2\rho}{(1+b^2)^2}
\end{eqnarray}
It is easy to see that in (\ref{decaytensor})
\begin{eqnarray}
T^{\mu\nu}&=&T^{\mu\nu}_{rt}+T^{\mu\nu}_\infty\nonumber\\
T^{\mu\nu}_{rt}&=&diag[\rho,T_s,T_s,T_s]\nonumber\\
T_s&=&\frac{\rho}{1+b^2}(1+b^2-e^2)[\tanh^2(\frac{\sqrt{1+b^2-e^2}\;t}{l_s\sqrt{k(1+b^2)}})-1]
\end{eqnarray}
This is a well-known phenomenon: when turning on an
electromagnetic field in the worldvolume of an unstable Dbrane,
the Dbrane decays away, leaving out the string fluid. So the
energy-momentum tensor is the sum of two parts: one is for the
decaying Dbrane as ${T}_{rt}$, which represents the asymptotic
pressureless matter; and the second part ${T}_\infty$ for the
string fluid which is independent of time.

Several remarks are in order: \\
(1) This bounded solution exists when the conserved energy is not
too large to make the Dbrane escape the attractive gravitational
force. Actually, from (\ref{decaytensor}), we notice that
\begin{eqnarray}\label{uperbound}
H\dot{r}^2(1+b^2)=1+b^2-e^2-\frac{\tau_p^2(1+b^2)^2}{H\rho^2},
\end{eqnarray}
Dbrane approaches its largest distance $r$ from NS5-branes at the
turning point $\dot{r}|_{r_{max}}=0$. We want our solution bounded
in the near throat region, that is $r_{max}\ll l_s\sqrt{k}$ or
$H(r_{max})\gg 1$. From (\ref{uperbound}), we obtain:
\begin{eqnarray}
H(r_{max})=\frac{\tau_p^2}{\rho^2}\frac{(1+b^2)^2}{1+b^2-e^2}\gg
1\nonumber\\
\Rightarrow \frac{\rho}{\tau_p}\ll\frac{1+b^2}{\sqrt{1+b^2-e^2}}
\end{eqnarray}
On the other hand, the string coupling should also be small in
order that our tree level analysis applies, which gives the bound
\begin{eqnarray}
\exp(\Phi)=A\cosh\frac{\sqrt{1+b^2-e^2}t}{l_s\sqrt{k(1+b^2)}}\ll1
\Rightarrow\frac{\rho}{\tau_p}\gg
\frac{g_s(1+b^2)}{\sqrt{1+b^2-e^2}}
\end{eqnarray}
So in the case
\begin{eqnarray}
\frac{g_s(1+b^2)}{\sqrt{1+b^2-e^2}}\ll\frac{\rho}{\tau_p}\ll\frac{1+b^2}{\sqrt{1+b^2-e^2}},
\end{eqnarray}
the near throat approximation applies, and there is a long period
of time in which the tree level analysis is good.\\
(2) For an observer on NS5-branes at asymptotic infinity, the
velocity of Dp-brane is
\begin{eqnarray}
\mathcal{V}&=\frac{dR}{dt}=&\frac{\sqrt{1+b^2-e^2}}{A\sqrt{1+b^2}}\frac{\tanh\frac{\sqrt{1+b^2-e^2}t}{l_s\sqrt{k(1+b^2)}}}
{\cosh\frac{\sqrt{1+b^2-e^2}t}{l_s\sqrt{k(1+b^2)}}}\nonumber\\
&\frac{\sim\sim}{t\rightarrow\infty}&\frac{1}{\cosh\frac{\sqrt{1+b^2-e^2}t}{l_s\sqrt{k(1+b^2)}}}.
\end{eqnarray}
From the NS5-branes point of view, the velocity of the Dbrane
slows down to zero when falling to the NS5-branes and the Dbrane
will get close to the NS5-branes in an infinite amount of time.
This is the
familiar redshift phenomenon in general relativity.\\
(3) On the other hand, the process seen by the observer on the
Dp-brane is quite different. Due to the existence of the
electromagnetic field strength, the open string dynamics should be
described by the open string metric
$\mathcal{G}_{\mu\nu}$\cite{NCOS}. The comoving time $\tau$ is
related to $t$ by $d\tau=\sqrt{-\mathcal{G}_{00}}dt$. Without loss
of generality , we turn off the magnetic field $b$ (as argued
above, we can boost to the rest frame of fundamental strings so as
to make b=0) and then
\begin{eqnarray}
\frac{d\tau}{dt}=\sqrt{1-e^2-(1-e^2)(\tanh^2\frac{\sqrt{1-e^2}\;t}{l_s\sqrt{k}})}
=\sqrt{(1-e^2)\frac{1}{\cosh^2\frac{\sqrt{1-e^2}\;t}{l_s\sqrt{k}}}}
\end{eqnarray}
which lead to
\begin{eqnarray}
\tan\frac{\tau}{2l_s\sqrt{k}}&=&e^{\frac{\sqrt{1-e^2}\;t}{l_s\sqrt{k}}}\nonumber\\
R&=&\frac{\sqrt{k}l_s}{A}\sin\frac{\tau}{l_s\sqrt{k}}.
\end{eqnarray}
In the above, $\tau$ runs from 0 to $\sqrt{k}l_s\pi$. Therefore,
from Dp-brane's point of view, it approaches the NS5-branes in a
finite amount of time, its velocity does not decay (of course, in
the very late time, our analysis is invalidated by the strong
string coupling). Actually, the Dbrane seems to feel no effect of
electric field strength since in terms of comoving time $\tau$,
the motion $R(\tau)$ is independent of $e$ except the amplitude of the oscillation. \\
(4) As usual, there exists a critical value for the electric field
$e^2\rightarrow1+b^2$, in which case the Lorentz factor
$\sqrt{(1+b^2-e^2)/(1+b^2)}$ vanishes. And when it approaches this
limit, the Dbrane stays still, since there is a balance between
the gravitational attraction and the electromagnetic repulsion.
More precisely, a good description of the Dbrane should be a NCOS theory\cite{NCOS}.\\
(5) Similarly, one may discuss the case  when one transverse
direction of the NS5-branes get compactified on a
$S^1$\cite{Kutasov3}. We label the compact direction by $y$ and
other three directions by $\vec{z}$. In the near throat region,
for the scalar fields $y$ and $\vec{z}$, the effective potential
is $(1-\frac{e^2}{1+b^2})h^{-1}$, where
\begin{equation}
h=\frac{k}{2r|\vec{z}|}\frac{\sinh(|\vec{z}|/r)}{\cosh(|\vec{z}|/r)-\cos(y/r)}.
\end{equation}
When $y=\pi r$ and $|\vec{z}|=0$, there is no force on the
Dp-brane. But this is an unstable configuration, the fluctuations
turn out to be of
\begin{equation}
m^2_y\sim -(1-\frac{e^2}{1+b^2})\frac{1}{k}, \hspace{5ex} m^2_z
\sim (1-\frac{e^2}{1+b^2})\frac{1}{k}.
\end{equation}
Therefore, one has a tachyonic mode around $y=\pi r$. This
unstable Dp-brane configuration reminds us of non-BPS branes. A
remarkable fact is that when $e$ gets close to the critical value,
due to the balance between the gravitational attraction and the
electromagnetic repulsion, the tachyon mode tends to be massless,
reminiscing of NCOS.

\section{Boundary states}

The conformal field theory in the background of (\ref{metric}) is
described by the CHS superconformal system in the near throat
limit. From (\ref{metric}) it is easy to see the topology near
horizon looks like $R^{1,5}\times R^1\times S^3$, and the
conformal field theory
factorizes accordingly. The three factors are: \\
$R^{1,5}$ : the free string theory targeted in six dimensional
Minkowski space or the NS5 brane world volume.\\
$R^1$ : a linear dilaton CFT with the dilaton proportional to
proper distance $\phi$ : $\Phi=-\frac{\phi}{\sqrt{k}l_s}$.\\
$S^3$ : the level $k$ $SU(2)$ supersymmetric WZW model.

The boundary states describing the Dbrane in the configuration
without an electromagnetic field were given in \cite{Nakayama}.
The $SU(2)_k$ WZW theory in general allows $k+1$ different
boundary states corresponding to BPS D-branes
\cite{Apikian},\cite{Nepomechie}:
\begin{eqnarray}\label{su2}
|B,l\rangle=\frac{1}{\sqrt{2}}\sum_{2j=0}^{k}(\frac{S^l_j}{\sqrt{S^0_j}}|j\rangle\rangle_{NSNS}+\frac{S'^l_j}{\sqrt{S'^0_j}}|j\rangle\rangle_{RR})\nonumber\\
|j\rangle\rangle_{NS,R}=\sum_{\psi\in\mathcal{H}_{NS,R}}e^{i\phi(j,\psi)}|j,\psi\rangle_L|j,\psi\rangle_R,
\end{eqnarray}
where the sum goes over all left-right symmetric states
constructed from $SU(2)$ prime $|j\rangle$.

 The space-like
world-volume of the D-brane is the standard one (we neglect the
ghost part through out):
\begin{equation}\label{Space}
|B_{space}\rangle=N_p\prod_{i=1}^{i=5}e^{\sum_n\frac{1}{n}\alpha^i_{-n}\bar{\alpha}^i_{-n}}
\end{equation}
Here $N_p$ is a normalization factor proportional to $T_p$. The
fermionic part is a little complicated and not essential. We refer
the reader to
\cite{Vecchia}.

The remaining two dimensions $(\phi,t)$ are related as the
on-shell motion of D-brane (\ref{EOM}) with vanishing $e$ and $b$.
This is described by the supersymmetric ``hairpin brane"
\cite{Lukyanov},\cite{Nakayama}:
\begin{eqnarray}\label{hairpin}
|B;P,Q\rangle^{\sigma}=\int_0^\infty dp\int_{-\infty}^\infty
d\omega \Psi^{\sigma}_{P,Q}(p,q)|p,\omega\rangle\rangle^{\sigma}
\end{eqnarray}
where $\sigma$ represents R-R sector or NS-NS sector sector,
$|p,\omega\rangle\rangle$ is the Ishibashi states associated to
some irreducible massive characters, and the wave functions are (as
in \cite{Nakayama}, we focus on the case $P=Q=0$)
\begin{eqnarray}\label{wavefunction}
\Psi^{(NS)}_{P=0,Q=0}(p,\omega)&=&\frac{-i\sqrt{2}\mathcal{Q}e^{i\frac{2p}{\mathcal{Q}}ln\tilde{r}}\sinh(\frac{2\pi
p}{\mathcal{Q}})}{2\cosh[\frac{\pi}{\mathcal{Q}}(p+\omega)]\cosh[\frac{\pi}{\mathcal{Q}}(p-\omega)]}\bullet
\frac{\Gamma(-i\mathcal{Q}p)\Gamma(1-i\frac{2p}{\mathcal{Q}})}{\Gamma(\frac{1}{2}+i\frac{\omega}{\mathcal{Q}}-i\frac{p}{\mathcal{Q}})\Gamma(\frac{1}{2}-i\frac{\omega}{\mathcal{Q}}-i\frac{p}{\mathcal{Q}})}\\
\Psi^{(NS)}_{P=0,Q=0}(\phi,t)&=&\frac{\sqrt{2}}{\pi\mathcal{Q}(2\cosh\frac{\mathcal{Q}t}{2})^{\frac{2}{\mathcal{Q}^2}+1}}exp[-\frac{\phi
'}{\mathcal{Q}}-\frac{e^-\frac{\phi'}{\mathcal{Q}}}{(2\cosh\frac{\mathcal{Q}t}{2})^{\frac{2}{\mathcal{Q}^2}}}]\,
,\phi'=\phi-\frac{2}{\mathcal{Q}}ln\tilde{r}\\
\Psi^{(R)}_{P=0,Q=0}(p,\omega)&=&\frac{-i\sqrt{2}\mathcal{Q}e^{i\frac{2p}{\mathcal{Q}}ln\tilde{r}}\sinh(\frac{2\pi
p}{\mathcal{Q}})}{\cosh\frac{2\pi
p}{\mathcal{Q}}-\cosh\frac{2\pi\omega}{\mathcal{Q}}}\bullet
\frac{\Gamma(-i\mathcal{Q}p)\Gamma(1-i\frac{2p}{\mathcal{Q}})}{\Gamma(1-i\frac{\omega}{\mathcal{Q}}-i\frac{p}{\mathcal{Q}})\Gamma(i\frac{\omega}{\mathcal{Q}}-i\frac{p}{\mathcal{Q}})}\\
\Psi^{(R)}_{P=0,Q=0}(\phi,t)&=&e^{\mathcal{Q}t}{2}\frac{\sqrt{2}}{\pi\mathcal{Q}(2\cosh\frac{\mathcal{Q}t}{2})^{\frac{2}{\mathcal{Q}^2}+1}}\exp[-\frac{\phi
'}{\mathcal{Q}}-\frac{e^-\frac{\phi'}{\mathcal{Q}}}{(2\cosh\frac{\mathcal{Q}t}{2})^{\frac{2}{\mathcal{Q}^2}}}]\,
,\phi'=\phi-\frac{2}{\mathcal{Q}}ln\tilde{r},
\end{eqnarray}
where $\mathcal{Q}=\frac{2}{l_s\sqrt{k}}\, ,
\tilde{r}=\frac{2}{A}\,$. Note that the NS sector wave function
and R sector wave function are related by the spectrum flow:
\begin{eqnarray}
\Psi^{(NS)}(p,\omega)&=&\Psi^{(R)}(p,\omega-i\frac{\mathcal{Q}}{2})\\
\Psi^{(NS)}_{P=0,Q=0}(\phi,t)&=&e^{-\mathcal{Q}t}{2}\Psi^{(R)}_{P=0,Q=0}(\phi,t)
\end{eqnarray}

Now we apply the T-duality approach developed in \cite{Vecchia},
\cite{Rey} to obtain the microscopic description of the Dbrane
with the electromagnetic field turned on. For simplicity, we set
$b=0$ first and perform a chain of maps, which
retains the solvability of the BCFT. \\
Step 1: In the absence of an electromagnetic field, compactify
$x_1$ on a circle of radius $R_1$, and T-dual along $x_1$. The
left and right moving parts of the world sheet scalar fields
$X^a(z,\bar z)$ are denoted by $X^a_L, X^a_R$. Under the T-duality
\begin{eqnarray}
\left(\begin{array}{c}
X_L^0\\
X_L^1\\
\end{array}\right)\mapsto
\left(\begin{array}{cc} +1&0\\
0&+1\\
\end{array}\right)
\left(\begin{array}{c}
X_L^0\\
X_L^1\\
\end{array}\right)\nonumber\\
\left(\begin{array}{c}
X_R^0\\
X_R^1\\
\end{array}\right)\mapsto
\left(\begin{array}{cc} +1&0\\
0&-1\\
\end{array}\right)
\left(\begin{array}{c}
X_R^0\\
X_R^1\\
\end{array}\right)
\end{eqnarray}
In this process, a Dp-brane turns into an array of
D(p-1)-branes on a circle of radius $\tilde{R_1}=2\pi/R_1$.
Take $\tilde{R_1}\rightarrow\infty$ limit, isolating a localized
D(p-1)-brane, the boundary state for $X^1$ is turned into a
Dirichlet one:
\begin{equation}\label{bound1}
|B,X^1\rangle=\exp(+\sum_{n=1}^{\infty}\frac{1}{n}\alpha^1_{-n}\bar{\alpha}^1_{-n})\delta(\hat{x}^1))|0\rangle
\end{equation}
Step 2: Boost the D(p-1)-brane along $x^1$ to a velocity $e$,
the world sheet fields are then transformed as
\begin{eqnarray}
\left(\begin{array}{c}
X_L^0\\
X_L^1\\
\end{array}\right)\mapsto \gamma
\left(\begin{array}{cc} 1&e\\
e&1\\
\end{array}\right)
\left(\begin{array}{c}
X_L^0\\
X_L^1\\
\end{array}\right)\nonumber\\
\left(\begin{array}{c}
X_R^0\\
X_R^1\\
\end{array}\right)\mapsto \gamma
\left(\begin{array}{cc} 1&e\\
e&1\\
\end{array}\right)
\left(\begin{array}{c}
X_R^0\\
X_R^1\\
\end{array}\right)
\end{eqnarray}
where $\gamma=1/\sqrt{1-e^2}$ is a Lorentz contraction factor.
After this operation, the oscillators in the boundary state
(\ref{bound1}) are transformed in the same way. But because of the constraint of
$\hat{x}^1$ in (\ref{bound1}) the zero mode transforms as:
\begin{eqnarray}\label{contract}
\hat{x}^0 &\mapsto& \gamma^{-1}x^0 \nonumber\\
\delta(\hat{x}^1) &\mapsto&
\gamma^{-1}\delta(\hat{x}^1+e\hat{x}^0)
\end{eqnarray}
It is clear that the boundary state of the hairpin is easily
transformed, since $\hat{x}^0$ is not mixed with $\hat{x}^1$,
just Lorentz contracts.

Step 3: T-dual along $X^1$ again, and we return to a Dp-brane.
Under the T-duality:
\begin{eqnarray}
\left(\begin{array}{c}
X_L^0\\
X_L^1\\
\end{array}\right)\mapsto
\left(\begin{array}{cc} +1&0\\
0&+1\\
\end{array}\right)
\left(\begin{array}{c}
X_L^0\\
X_L^1\\
\end{array}\right)\nonumber\\
\left(\begin{array}{c}
X_R^0\\
X_R^1\\
\end{array}\right)\mapsto
\left(\begin{array}{cc} +1&0\\
0&-1\\
\end{array}\right)
\left(\begin{array}{c}
X_R^0\\
X_R^1\\
\end{array}\right).
\end{eqnarray}
Moreover, the zero mode part of the boundary state transforms as:
\begin{eqnarray}
\gamma^{-1}\delta(\hat{x}^1+e\hat{x}^0)=\gamma^{-1}\sum_{n\in\mathcal{Z}}e^{i\frac{n}{\tilde{R_1}}(\hat{x}^1+e\hat{x}^0)}|0\rangle
\mapsto \gamma^{-1}\sum_{m\in\mathcal{Z}}e^{i m
R_1(\hat{x}^1_L-\hat{x}^1_R+e\hat{x}^0)}|0\rangle
\end{eqnarray}
In the decompactification  limit, $R_1\rightarrow\infty$, only $m=0$
term contributes, yielding a Born-Infeld factor to the boundary
state: $\sqrt{1-e^2}|0\rangle$. After this operation, we come back
to a Dp-brane but with a constant electric field $F_{01}=e$
turned on. Also, the timelike direction is lorentz contracted as
in (\ref{contract}).

When turning on both the electric and magnetic fields, we have:
\begin{eqnarray}
|0\rangle&\mapsto&\tilde{\gamma}^{-1}\gamma^{-1}|0\rangle\label{vacuumrescale} \\
t&\mapsto&\gamma^{-1}x^0\label{scalet}\\
\left(\begin{array}{c} \alpha^0_{-n}\\
\alpha^1_{-n}\\
\alpha^2_{-n}\end{array}\right) &\mapsto&\Lambda^{-1}\Omega^{-1}
\left(\begin{array}{c}\alpha^0_{-n}\\
\alpha^1_{-n}\\
\alpha^2_{-n}\end{array}\right)\\
\left(\begin{array}{c} \bar{\alpha}^0_{-n}\\
\bar{\alpha}^1_{-n}\\
\bar{\alpha}^2_{-n}\end{array}\right)&\mapsto&\Lambda\Omega
\left(\begin{array}{c}\bar{\alpha}^0_{-n}\\
\bar{\alpha}^1_{-n}\\
\bar{\alpha}^2_{-n}\end{array}\right)\\
\left(\begin{array}{c} \psi^0_{-n}\\
\psi^1_{-n}\\
\psi^2_{-n}\end{array}\right)&\mapsto&\Lambda^{-1}\Omega^{-1}
\left(\begin{array}{c}\psi^0_{-n}\\
\psi^1_{-n}\\
\psi^2_{-n}\end{array}\right)\\
\left(\begin{array}{c} \bar{\psi}^0_{-n}\\
\bar{\psi}^1_{-n}\\
\bar{\psi}^2_{-n}\end{array}\right)&\mapsto&\Lambda\Omega
\left(\begin{array}{c}\bar{\psi}^0_{-n}\\
\bar{\psi}^1_{-n}\\
\bar{\psi}^2_{-n}\end{array}\right)
\end{eqnarray}
where
\begin{eqnarray}
\Lambda=\left(\begin{array}{ccc} \gamma&0&\gamma e'\\
0&1&0\\
\gamma e'&0&\gamma\end{array}\right)\, ,\hspace{5ex}\Omega=\left(\begin{array}{ccc} 1&0&0\\
0&\tilde{\gamma}&-\tilde{\gamma}b\\
0&\tilde{\gamma}b&\tilde{\gamma}\end{array}\right)\nonumber\\
\gamma=\frac{1}{\sqrt{1-e'^2}},\hspace{3ex}\tilde{\gamma}=\frac{1}{\sqrt{1+b^2}},\hspace{3ex}e'=\frac{e}{\sqrt{1+b^2}}.
\end{eqnarray}
It is not hard to see that the modified hairpin solution
(\ref{EOM}) is consistent with the transformation (\ref{scalet}).
Therefore,
we propose here that when the electromagnetic field is turned on,
the hairpin brane gets modified as in (\ref{EOM}),  and the
boundary state can be obtained in a simple way: the Ishibashi
states are modified only by the rescaling (\ref{scalet}), and the
boundary wave function
$\Psi^{\sigma}_{P,Q}(p,\omega)\mapsto\Psi^{\sigma}_{P,Q}(p,\gamma
\omega)$.\footnote{Here p is the momentum in the $R_\phi$
direction as in \cite{Sahakyan}.} One subtle point is that we are
short of a rigorous proof that the modified hairpin (\ref{EOM}) is
a marginal deformation. Nevertheless, the experience on the
rolling tachyon tells us that the T-duality approach gives us the
correct boundary states with the electromagnetic fields turning
on. We believe that the modified hairpin is a truly marginal
operator and our proposal is correct.

The same Fourier transformation procedure as in \cite{Nakayama}
leads to the coordinate space wave function which peaks at the
trajectory (\ref{EOM}). This is a consistent check of our
proposal.

The boundary states for other space-like directions without the
electromagnetic field remain the same, while the boundary states
for ${X^{1,2},\psi^{1,2}}$ now are written as:
\begin{equation}\label{Spaceshifed}
|B,X^i,i=1,2\rangle=N_pe^{\sum_n\frac{1}{n}\alpha'^i_{-n}\bar{\alpha'}^i_{-n}}\,\,\,
, \alpha'^i_{-n}=[(\Omega\Lambda)^{-1}]^{i}_{j}\alpha^j\, \,\,,
\bar{\alpha'}^i_{-n}=(\Lambda\Omega)^{i}_{j}\bar{\alpha}^j,
\end{equation}
with the fermionic part transforming in the same way. There are
factors $\gamma,\tilde{\gamma}$  in the oscillator part, we can
choose a gauge in which the contribution to the total energy and
particle number of closed string radiation comes only from
timelike zero modes \cite{Liu}, so that only the vacuum rescaling
in
(\ref{vacuumrescale}) is relevant.  \\
Now we are ready to calculate the closed string emission rate from
the above boundary states.

\section{Closed string radiation: without winding strings}

Note first that the amplitude for the emission of a closed string
mode $V$ is
\begin{equation}
\mathcal{A}\sim\frac{\langle V|B\rangle}{\sqrt{E}},
\end{equation}
where $\langle V|B\rangle$ is the one-point function of a closed
string vertex operator. One should also keep in mind that only the
left-right symmetric closed string states are contained in the
spectrum of a Dbrane. In order to calculate the above one-point
function, one needs a full description of the boundary state. The
boundary state factorizes into three independent pieces:
\begin{itemize}
\item The hairpin part is already given above, and its
contribution is determined by the wave function;
\item The boundary state of the spacelike directions without the electromagnetic field
is the same as the standard one, while the one of the directions
involving the electromagnetic field get changed to
(\ref{Spaceshifed});
\item As shown in (\ref{su2}), the $SU(2) WZW$ part contributes a phase
factor as well as an overall constant factor $S^i_jS'^i_j$.
Without loss of generality, we take $l=0$ in (\ref{su2}), that is a
Dbrane localized in $S^3$ (the group manifold of $SU(2)$).
\end{itemize}

Now the cross-section is
\begin{eqnarray}
|\langle
V(j,p,\omega)|B\rangle|^2&\simeq&\gamma^{-2}\tilde{\gamma}^{-2}\frac{\pi}{k}\frac{\sinh(\pi\sqrt{k}p)}{(\cosh\pi\sqrt{k}\gamma\omega+
\cosh\pi\sqrt{k}p)\sinh\frac{\pi p}{\sqrt{k}}}S^j_0\nonumber\\
\omega&=&\sqrt{p^2+k_\perp^2+4n+\frac{j(j+1)}{k}}
\end{eqnarray}
where $p$ is the momentum in the R (linear dilaton) direction,
$\vec{k}_\perp$ is the momentum parallel to the NS5-branes  and
transverse to the Dbrane, $n$ is the total oscillator level and
$j$ is the $SU(2)$ spin of the primary state we consider here.

We are interested in the total average number  and energy of
particles emitted by the Dbrane, which are given by
 \bea
 \frac{{\mathcal N}}{V}&=& \sum_{\mbox{states}}\frac{1}{2\omega}|\langle
V(j,p,\omega)|B\rangle|^2 \\
 \frac{{\mathcal E}}{V}&=& \sum_{\mbox{states}}|\langle
V(j,p,\omega)|B\rangle|^2 \eea
  From the form
of the above cross-section, the possible divergence may come from
the large $p$ region. Thus $\frac{j(j+1)}{k}$ in $\omega$ can be
neglected in this region. Recall that the density of left-right
symmetric closed string states for large level $n$ is $d_n\sim
n^{-\frac{q+1}{2}}e^{\pi\sqrt{4n}\sqrt{\frac{2k-1}{k}}}$, where
$2q=6$ is the number of non-compact spatial directions. So another
possible divergence could be from the exponential growth of the
states with large level. Set $l^2=4n$, and change the summation
over $n$ into the integral over $l$,  the average number of total
emitted particle could be approximated by:
\begin{equation}
\frac{{\mathcal N}}{V}\sim\gamma^{-2}\tilde{\gamma}^{-2}\int
d\theta d\phi
dr(\cos\theta)^{d-1}(\sin\theta)^{1-q}(\sin\phi)^{-q}r^{d-q}\exp[-\pi
r(\gamma-\sin\theta((1-\frac{1}{k})\cos\phi+\sqrt{\frac{2k-1}{k^2}}\sin\phi))]
\end{equation}
after introducing a set of spherical coordinates
\begin{eqnarray}
r&=&\sqrt{k}\omega,\nonumber\\
p\sqrt{k}&=&r\sin\theta\cos\phi,\nonumber\\
l\sqrt{k}&=&r\sin\theta\sin\phi,\nonumber\\
k_\perp\sqrt{k}&=&r\cos\theta.
\end{eqnarray}
Here $d$ is the number of flat space-like directions transverse to
the Dbrane in NS5-branes\footnote{Here for Dp-brane, d=5-p.}. The
expression in the exponential in the integrand could be written as
\begin{eqnarray}
-\pi
r(\gamma-\sin\theta((1-\frac{1}{k})\cos\phi+\sqrt{\frac{2k-1}{k^2}}\sin\phi))\nonumber\\
=-\pi r(\gamma-\sin\theta \cos(\phi-\chi))
\end{eqnarray}
where
 \bea\label{defchi}
 \cos\chi=1-\frac{1}{k}, \hspace{5ex}
 \sin\chi=\sqrt{\frac{2k-1}{k^2}}.
 \eea

Since $\gamma>1$ when $e\neq0$, the integral is always
finite and the steepest descent approximation gives:
\begin{eqnarray}
\frac{{\mathcal N}}{V}&\sim&\gamma^{-2}\tilde{\gamma}^{-2}\int
dr\int dx r^{d-q}x^{d-1}e^{-\pi
r(\gamma-1+\frac{1}{2}x^2)} \nonumber\\
 &\sim&\gamma^{-2}\tilde{\gamma}^{-2}\int dr e^{-\pi r(\gamma-1)}
r^{d/2-q}\nonumber \\
&\sim&\gamma^{-2}\tilde{\gamma}^{-2}(\gamma-1)^{\frac{4-d}{2}},
\end{eqnarray}
Similarly, one can obtain the total emitted energy
\begin{eqnarray}
\frac{\mathcal{E}}{V}\sim\gamma^{-2}\tilde{\gamma}^{-2}\int dr
e^{-\pi r(\gamma-1)}r^{d/2-q+1}
&\sim&\gamma^{-2}\tilde{\gamma}^{-2}(\gamma-1)^{\frac{2-d}{2}}.
\end{eqnarray}

It is obvious that when \\
(1) $e\neq0$ (especially this is independent of the magnetic
field), the above formula is finite thus indicates the validity of
our tree
level analysis.\\
(2) $e\rightarrow0$, that is $\gamma\rightarrow 1$. For the
Dp-brane, $d=5-p$, when $p> 3$ the total emitted energy is finite
but for $p\leq 3$ it is infinite, implying the breakdown of the
tree level
treatment.\\
(3) $e$ tends to the critical value $e=1+b^2$, $\gamma\rightarrow
\infty$, the emitted energy
$\mathcal{E}\sim\tilde{\gamma}^{-2}\gamma^{\frac{-d}{2}}\rightarrow
0 $. This means that Dp-brane decouples from the closed string.

\section{Closed string radiation: with winding strings}

The above result shows that the total energy of the closed string
radiation is finite with the presence of nonvanishing electric
field. This is similar to what happens in the rolling tachyon case
\cite{Nagami}. However, the authors of \cite{Yi2} found that in
the rolling tachyon case, taking the winding closed string into
account the radiation is UV divergent when $p\leq 3$, and as a
result the emitted winding closed string will dominate the tachyon
matter.

For simplicity, we take $b=0$ (so
$\tilde{\gamma}=1,\gamma=\frac{1}{\sqrt{1-e^2}})$. Let us
compactify $x^1$ direction on a circle of radius R, and wrap the
Dbrane on this circle. The boundary state for zero mode is
\begin{eqnarray}\label{wind}
|B,m\rangle^{\sigma}= \sqrt{1-e^2} \sum_m\int_0^\infty
dp\int_{-\infty}^\infty d\omega
\Psi^{\sigma}(p,\omega)\cdot \nonumber\\
|p,\omega_s=p^0_L=p^0_R=\sqrt{1-e^2}\omega+eRm,
p^1_L=-p^1_R=Rm\rangle\rangle^{\sigma}
\end{eqnarray}
where $m$ is the winding number. Now for a left-right symmetric
closed string with winding number $m$ and energy $\omega$, the
cross section is
\begin{eqnarray}
|\langle
V(j,p,m,\omega_s)|B\rangle|^2&\simeq&\gamma^{-2}\frac{\pi^2}{k}\frac{\sinh(\pi\sqrt{k}p)}
{(\cosh\pi\sqrt{k}\gamma(\omega-eRm)+\cosh\pi\sqrt{k}p)\sinh\frac{\pi p}{\sqrt{k}}}S^j_0\nonumber\\
\omega_s&=&\sqrt{p^2+k_\perp^2+4n+m^2R^2+\frac{j(j+1)}{k}},
\end{eqnarray}
and the average number and energy of total emitted particles are
\bea
 \frac{{\mathcal N}}{V}&=& \sum_{\mbox{states}}\frac{1}{2\omega_s}|\langle
V(j,p,m,\omega_s)|B\rangle|^2 \\
 \frac{{\mathcal E}}{V}&=& \sum_{\mbox{states}}|\langle
V(j,p,m,\omega_s)|B\rangle|^2.
 \eea
 Note that now the summation over states include not only all
 states of different levels but also of different windings.

 As in the above section, we are interested in the high
energy region so we can neglect the term $j(j+1)/k$ in $\omega_s$.
The total emitted energy is
\begin{eqnarray}
\frac{\mathcal{E}}{V}\sim R(1-e^2)\sum_m\int dp dl d^dk_\perp
l^{-q}e^{l\pi\sqrt{\frac{2k-1}{k}}}\cdot\nonumber\\
e^{-\pi\sqrt{k}\gamma(\sqrt{p^2+k_\perp^2+l^2+m^2R^2+\frac{j(j+1)}{k}}-eRm)+\pi\sqrt{k}p-\frac{p\pi}{\sqrt{k}}
}
\end{eqnarray}
where $l^2=4n$. In the large $n$ limit, we can replace the sum
over $m$ to the integral. Introduce the spherical coordinates
\begin{eqnarray}
r&=&\sqrt{k}\sqrt{(mR)^2+p^2+l^2+k^2_\perp} \nonumber \\
mR\sqrt{k}&=& r\cos\theta \nonumber \\
k_\perp\sqrt{k}&=&r\sin\theta\cos\varphi \nonumber \\
l\sqrt{k}&=&r\sin\theta\sin\varphi\cos\phi \nonumber \\
p\sqrt{k}&=&r\sin\theta\sin\varphi\sin\phi,
\end{eqnarray}
the integral becomes
 \bea
 \frac{\mathcal{E}}{V} &\sim & \int dr d\theta d\varphi d\phi
 r^{d-q+2}(\sin\theta)^{d+1-q}(\sin\varphi)^{1-q}(\cos\varphi)^{d-1}(\cos\phi)^{-1}\times
 \nonumber \\
  & & \exp(-\pi
  r(\gamma(1-e\cos\theta)-\sin\theta\sin\varphi\sin(\phi-\chi)))
  \eea
where $\chi$ is defined in (\ref{defchi}).

The factor in the exponential can be written as
 \bea
 -\pi r
 \gamma(1-\cos\delta\cos\theta-\sin\delta\sin\theta\sin\varphi\sin(\phi-\chi)),
 \eea
where
 \bea
 \cos\delta=e, \hspace{5ex} \sin\delta=\sqrt{1-e^2}.
 \eea

Therefore, we see that the integral has power-like divergence in
$r$ when $\phi\sim \chi+\frac{\pi}{2}$, $\varphi\sim
\frac{\pi}{2}$ and $\theta\sim \delta$. This contrasts with the
unwinding case, where $\theta=\pi/2$ and the above integral is
convergent. By using the steepest descent method, we find
 \bea
 \frac{\mathcal{E}}{V} \sim \int dr r^{\frac{d}{2}-q+2},
 \eea
 which is always divergent.
 Similarly, one can calculate the average number of total emitted particles
 \bea
 \frac{\mathcal{N}}{V} \sim \int dr r^{\frac{d}{2}-q+1}
 \eea
Therefore, taking into account the wound strings, the total energy
of emitted string is divergent. This suggests that the tachyon
matter consists mainly of the highly wound strings.

There are two interesting limits in this case. When $e=0$, namely
turning off all the electromagnetic field, the dominant
contribution in the exponential is from $\theta\simeq \delta=
\pi/2$ and the integration by steepest descent method tells us
that the total emitted energy is still divergent. On the other
hand, when $e$ reach the critical value, $\gamma \rightarrow
\infty$ and it is not hard to see that the total emitted energy
vanish.

\section{Conclusion}

In this paper, we investigated the dynamics of a Dp-brane probe
near a stack of NS5-branes with constant electromagnetic field
turned on. The result is similar to but differs in detail from the
rolling tachyon case: Dbrane falls  onto the NS5-branes with a
speed approaching a maximum velocity less than the speed of light
(as seen by the near throat observer), radiating closed string
modes, leaving out the string fluid on the NS5-branes. The whole
process is slowed by a Lorentz factor $\gamma>1$. The average
number and energy of the emitted particle  are both finite when
electric field $e$ and magnetic field $b$ satisfy
$0<e<\sqrt{1+b^2}$. When compactify the Dbrane on a circle and
take into account the winding string radiation, we find that the
total radiation always diverges, implying that the winding string
emission  dominates the tachyon matter. We also study several
properties of the system when $e$ approaches the critical value
$\sqrt{1+b^2}$, a limit related to NCOS theory, and this deserves
further study.

A more geometric system was constructed in \cite{Kutasov2} with
one transverse direction compactified. Our DBI analysis could be
easily generalized to that case, but the conformal field theory of
the system turns to be quite complicated so we are not able to
construct the boundary state of the Dbrane. We leave this problem
as a future plan.

Another interesting question is to analyze the modified hairpin
brane (\ref{EOM}) in the context of the conformal field theory. It
would be nice to prove
 it is truly marginal and construct its
boundary state directly from BCFT.

\section*{Acknowledgements}

The work of ML and BS was supported by a grant of NFSC and a grant
of Chinese Academy of Sciences, the work of BC was supported by a
grant of Chinese Academy of Sciences.

\bibliographystyle{amsplain}

\end{document}